\documentstyle[preprint,aps]{revtex}

\begin{document}
\draft
\preprint{}
\title{$c$-axis penetration depth of Hg-1201 single crystals}
\author{J.R. Kirtley}
\address{ IBM T.J. Watson Research Center, P.O. Box 218,
Yorktown Heights, NY 10598}
\author{K.A. Moler}
\address{ Department of Physics, Princeton University, Princeton,
NJ 08544}
\author{G. Villard and A. Maignan}
\address{CRISMAT-ISMRA, 6 Boulevard du Mar\'{e}chal Juin,
14050 Caen Cedex, France}
\date{\today}
\maketitle
\begin{abstract} 
We have magnetically imaged interlayer Josephson 
vortices emerging from an $ac$ face of single crystals of the single 
layer cuprate high-$T_c$ superconductor (Hg,Cu)Ba$_2$CuO$_{4+\delta}$.  
These images provide a direct measurement of the $c$-axis penetration 
depth, $\lambda_c \sim$ 10 $\mu$m.  This length is a factor of 10 
longer than predicted by the interlayer tunneling model for the 
mechanism of superconductivity in layered compounds, indicating that 
the condensation energy available through this mechanism is 100 times 
smaller than is required for superconductivity.  
\end{abstract}

\pagebreak
\narrowtext
In the interlayer tunneling (ILT) model for superconductivity in 
layered superconductors such as the 
cuprates\cite{wheatley,anderson1,chakravarty,anderson2}, transport of 
carriers between the planes is incoherent in the normal state, but 
coherent interlayer transport is allowed for Cooper pairs.  The 
coherent pair tunneling lowers the $c$-axis kinetic energy, supplying 
the superconducting condensation energy, $E_{c}$.  Anderson 
\cite{anderson2,andsci98}, Leggett\cite{leggett,legsci98}, and 
Chakravarty\cite{chakravarty2} have each argued that the comparison of 
the experimentally measured $c$-axis penetration depth, $\lambda_{c}$, 
and the value determined within the ILT model from the condensation 
energy, $\lambda_{ILT}$, is an important test of the ILT 
mechanism.  In all presently published versions of the theory, 
$\lambda_{c} \approx \lambda_{ILT}$.

The calculation of $\lambda_{ILT}$ and its comparison with experiment 
are most straightforward in materials with a single copper oxide layer 
per unit cell, such as La$_{2-x}$Sr$_x$CuO$_{4+\delta}$ 
($\lambda_{ILT} \simeq 3 \mu$m\cite{andsci98}), 
HgBa$_2$CuO$_{4+\delta}$ ($\lambda_{ILT} \simeq 1 
\mu$m\cite{andsci98}), and Tl$_2$Ba$_2$CuO$_{6+\delta}$ 
($\lambda_{ILT} \simeq 1 \mu$m\cite{andsci98}).  Different 
experiments on these materials have not provided a single answer to 
the question\cite{andsci98,legsci98}.  In 
La$_{2-x}$Sr$_x$CuO$_{4+\delta}$ (La-214), measurements of the 
Josephson plasma frequency $\omega_p = c 
\lambda_c^{-1}\epsilon^{-1/2}$ (where $c$ is the speed of light and 
$\epsilon$ is the dielectric constant of the interlayer 
medium)\cite{uchida,marel} are in good agreement with the predictions 
of the ILT model\cite{andsci98}.  In Tl$_2$Ba$_2$CuO$_{6+\delta}$ 
(Tl-2201), Moler {\it et al.} observed interlayer Josephson vortices 
with $\lambda_{c} = 17-21$ microns \cite{kamsci98}.  Measurements of 
the plasma resonance $\omega_{p}$ and dielectric constant $\epsilon$ 
by Tsvetkov {\it et al.} indicate $\lambda_{c} = 17$ 
microns\cite{marel,tsvetkov}.  In HgBa$_2$CuO$_{4+\delta}$ (Hg-1201), 
Panagopoulos {\it et al.} obtained a value of 
$\lambda_c(T=0)=1.36\pm0.16\mu$m from magnetic susceptibility data on 
oriented 
powders\cite{panagopoulos}.
As both Anderson\cite{andsci98} and 
Leggett\cite{legsci98} have pointed out, a confirmation of the 
$c$-axis penetration depth in Hg-1201 seems essential to resolving this 
issue.

In this Letter we
directly measure the $c$-axis penetration depth in Hg-1201 by magnetically
imaging interlayer Josephson vortices emerging from the ac
face of single crystals.
These measurements give $\lambda_c \simeq 10 \mu m$,
much longer than the value $\lambda_c \simeq 1 \mu m$
reported by Panagopoulos {\it et al.}\cite{panagopoulos}.
Our value for $\lambda_{c}$, when combined
with previous results from Tl-2201, make it appear unlikely that
presently published versions of the ILT model are the
correct description for superconductivity in the cuprate high-T$_c$
superconductors.

The Hg-1201 crystal growth has been described previously
\cite{pelloquin}.
Our method consists in preparing a Ba/Cu/O precursor in flowing oxygen,
and then mixing it with HgO to make
a 0.8:2:1.2 ratio of Hg:Ba:Cu. An alumina crucible containing
this powder is then
sealed in a silica tube. After 48 hours of thermal treatment, black,
platelet-like crystals
are extracted. The largest crystals (typical
dimensions $\sim$1$\times$1$\times$0.08mm)
are selected for transport and SQUID imaging measurements. EDX and electron
diffraction measurements lead to the formula Hg$_{0.8}$Cu$_{0.2}$ for the 
mixed
mercury layer. The copper position is displaced with respect to the Hg site
in the mixed layer as shown from structural refinements. No intergrowth nor
extended defects have been shown by high resolution 
microscopy\cite{pelloquin2}.
The Hg-1201 single crystals were mounted in epoxy so
that the ac face was aligned vertically, and then the sample
and epoxy were polished to make a flat, smooth surface to scan.

The magnetic imaging measurements were made with a scanning
SQUID microscope\cite{ssmapl}, in which a sample is scanned relative to
a superconducting pickup loop oriented nearly parallel to the sample
surface. The data are represented as the magnetic flux $\Phi_s$, the 
integral
of the normal component of the magnetic field over the pickup loop,
vs the position of the pickup loop in the $x,y$ plane.
The pickup loop is fabricated with
well-shielded leads to an integrated Nb-Al$_2$O$_3$-Nb SQUID. For these
measurements, we used a SQUID with our smallest pickup loop,
an octagon $L=4\mu$m in diameter
with 0.8$\mu$m linewidth.  The silicon substrate carrying the
SQUID was polished to a sharp tip a few microns from
the edge of the pickup loop, the substrate was oriented an angle
$\theta \sim 20^{\circ}$ from parallel to the sample, and scanned with
the tip in direct contact with the sample. Fits
to data on Abrikosov vortices indicate that the pickup loop is about 
$z_0$=1$\mu$m
above the surface of the sample under the experimental conditions
used here. Our images were made with both sample and SQUID
immersed in liquid helium at 4.2K.

Figure 2 shows a SSM image of a 512$\mu$m$\times$256$\mu$m area of
the crystal and surrounding epoxy. The outer edges of the crystal face
are indicated by dashed lines. The sample was cooled and imaged in
a field of about 10mG. Visible in this image are about 50 interlayer
Josephson vortices. These vortices are remarkably uniform in shape.
Some are sufficiently isolated
from their neighbors that overlapping fields do not interfere with
detailed modelling.
Although one must consider that
the vortices may be pinned in areas with unusually weakly coupled
planes, the uniformity of the vortex shapes over a large area of the
crystal face makes us believe that we are measuring an intrinsic 
penetration
depth. This belief is supported by the good agreement between bulk
plasma resonance measurements of $\omega_j$ and local interlayer Josephson 
vortex
imaging measurements of $\lambda_c$ in Tl-2201\cite{tsvetkov}.
Three vortices
chosen for further analysis are shown in figure 2. Cross-sections
through the image data parallel to the layers are displayed in Figure 3.

The decay of the observed vortex magnetic flux perpendicular
to the layers (along the c-axis) is
determined by the size of the pickup loop. In contrast, the extent of 
the images of the vortices along the layers provides a
direct measure of $\lambda_c$.

For quantitative modelling of the shape of the vortex we use
the results of Clem and Coffey\cite{clem}. Neglecting
the influence of the surface on the fields, the z-component of
the magnetic field of an interlayer vortex is given by\cite{clem}:
\begin{equation} B_z(x,y,z=0) = \frac{\Phi_0}{2 \pi \lambda_{ab}
\lambda_c} K_0(\tilde R),
\end{equation}
where $\lambda_{ab}$ is the in-plane penetration depth,
$K_0$ is a modified
Bessel function of the second kind of order 0, $\Phi_0=hc/2e$ is the
superconducting flux quantum, $h$ is Planck's constant,
$e$ is the charge on the electron, ${\tilde R} = ((s/2\lambda_{ab})^2 +
(x/\lambda_{ab})^2 + (y/\lambda_c)^2)^{1/2}$, $s$ is the interplane 
spacing, $x$ is the distance perpendicular to the planes, and $y$ is 
the distance parallel to the planes.
Since for our experiments 
$s << \lambda_{ab} << (L,z_0) << \lambda_{c}$,
we neglect both $s$ and $\lambda_{ab}$.
With this assumption, the fields from Eq. 1
are propagated to a height $z=z_0$\cite{blnkt}
and then summed over
the geometry of the pickup loop.  This model has
two free parameters:  $\lambda_c$, which determines the length of the 
vortex, and $z_0$, which determines the magnetic amplitude
of the vortex image.  Fits to the three cross-sections (Figure 3) yield
consistent values for the interlayer penetration depth
$\lambda_c = 10 \pm 1 \mu m$.

This measured value is about ten times longer than the theoretical 
value, $\lambda_{ILT}=1 \pm 0.5 \mu m$\cite{andsci98}.  Since the ILT 
condensation energy is proportional to $1/\lambda_{c}^{2}$, the ILT 
supplied condensation energy is about 100 times smaller than the 
estimated actual energy in Hg-1201\cite{andsci98}.

A more conventional estimate of $\lambda_c$ comes from the 
Lawrence-Doniach model\cite{lawrence}.  For diffusive pair transfer 
(parallel momentum not conserved) between superconducting 
layers\cite{lawrence,bulaevski,baratoff}, the Josephson current 
between two identical superconducting sheets at T=0 is given 
by\cite{baratoff} 
\begin{equation} J_0 = \frac{\pi \Delta(0)}{2 e 
R_{c,n}} \end{equation} 
where $\Delta(0)$ is the zero temperature 
energy gap and $R_{c,n}$ is the normal state $c$-axis interplane 
resistance.  The interlayer penetration depth is given by 
$\lambda_{\perp} = (c \Phi_0/8 \pi^2 s J_0)^{1/2}$\cite{clem}.  The 
application of this model requires at least two assumptions: that 
$R_n$ is temperature independent below $T_c$, or at least that the 
temperature dependence can be understood well enough for 
extrapolation\cite{chakravarty2}, and that the gap is 
$s$-wave\cite{sauls,levin,hirschfeld}.  $d$-wave superconductivity 
would tend to increase $\lambda_c$ from this value: in a purely 
tetragonal $d$-wave superconductor, with purely diffusive pair 
transfer, the coupling would be reduced to 
zero\cite{sauls,levin,hirschfeld}.  With these considerations in mind, 
one would not expect the Ambegaoker-Baratoff model to apply 
quantitatively, but it is nevertheless natural to look for a 
correlation between $\lambda_{c}$ and $R_{c,n}$.

Figure 4a shows measurements of the $c$-axis resistivity
of two of our Hg-1201 crystals as a function of temperature.
These measurements were made by
evaporating four Ag stripes, two on each
$ab$ face of the crystal, as diagrammed
in the inset of Fig 4a\cite{hardy}.
The contacts as prepared have high resistance, but after an annealing step
at 400$^o$C for 10 minutes the contact resistances drop to 1 to 2 Ohms.
We have carefully checked that such annealing does not alter the
superconducting transition by measuring the susceptibility of control
crystals before and after such an annealing step. Because of the relatively
high anisotropy of Hg-1201 and large thickness of our crystals ($\sim
80 \mu m$), these crystals are electrically thick\cite{montgomery}.
Therefore difficulties in separating the in- and out-of-plane 
conductivities
such as reported by Hussey {\it et al.}\cite{hussey} for
YBa$_2$Cu$_4$O$_8$ should not occur. The midpoint T$_c$=94K
of the $\rho_c(T)$ curves is consistent with the onset $T_c$=95-97K
observed magnetically on such crystals and confirms that the doping is
close to optimal. Taking $\rho_{c,n}$=$\sigma_{c,n}^{-1}$ = 0.5 Ohm-cm,
and a BCS gap value
2$\Delta$=3.54k$_B$T$_c$= 28 meV, leads to a predicted penetration depth
$\lambda_c = (\hbar c^2/4 \pi^2 \Delta \sigma_{c,n})^{1/2}$ =
8$\mu$m. 

Basov {\it et al.} \cite{basov}
have previously noted a correlation between the values of the
penetration depth $\lambda_c$ and the far-infrared
$c$-axis conductivity, $\sigma_{c,FIR}$. In Figure 4b
add our results for Hg-1201 and
Tl-2201\cite{kamsci98}, replacing $\sigma_{c,FIR}$ with $\sigma_{c,n}$
just above $T_c$.
The dashed line is Eq. (2),
assuming $2\Delta$ = 28meV.
This simple model, although it assumes the
same $\Delta$ for all cuprates, and has the shortcomings outlined above,
agrees with all of the measurements to within a factor of three.

It is not clear why our results for $\lambda_c$ are nearly a factor of 
7 larger than those of Panagopoulos {\it et al.}.  There are possible 
systematic errors in both measurements.  In the analysis of the 
imaging experiments, 
the spreading of the vortex near the surface is neglected.  This 
neglect is justified within our experimental resolution 
on the basis of a consideration of the free 
energy associated with vortex spreading in highly anisotropic 
superconductors\cite{huse}, as well as by the 
quantitative experimental agreement between $\lambda_{c}$ and 
$\omega_{p}$ in Tl-2201\cite{tsvetkov}.  For the powder magnetization 
results, the analysis depends on a very accurate characterization of 
the distribution of particle sizes; it is possible that this analysis 
could be skewed by the presence of a slightly higher number of large 
particles than used in the modelling\cite{panagpc}.  A more likely 
explanation is that the different samples do in fact have different 
$\lambda_{c}$ values.  First, the doping in the two samples is not 
identical.  It has been shown that the transport anisotropy of several 
materials can depend strongly on doping\cite{anisvsdop}, but it has 
also been shown that the anisotropy in Hg-1201 is nearly constant near 
optimal doping\cite{anisopt}.  Another possibility is the influence of 
the excess copper on the mercury layer.  Neutron studies on 
ceramics\cite{wagner,asab} conclude they also have mixed Hg/Cu layers.  
It is not clear whether the ceramics of Panagopoulos {\it et al.} have 
similar substitutions on the mercury layer.

If the Hg-1201 crystals used in this study and the Tl-2201 crystals
used in the previous study are true single-layer materials, then the
long $c$-axis penetration depths $\lambda_{c} \approx 10$ microns for
Hg-1201 and $\lambda_{c} \approx 20$ microns for Tl-2201 pose a serious
challenge to the ILT model.  Although one can hypothesize structural
explanations, perhaps based on the excess copper, to make these
results consistent with the ILT model, such explanations must be
found for both
(Hg,Cu)Ba$_2$CuO$_{4+\delta}$ and Tl$_2$Ba$_2$CuO$_{6+\delta}$.

These two results fall into a previously unfilled range of penetration
depths in the Basov correlation.  This correlation,
$\lambda_c \approx (\hbar c^2/4 \pi^2 \Delta \sigma_{c,n})^{1/2}$, works
surprisingly well given the simplicity of the theoretical model and
considering that details of the gap anisotropy, the band structure,
the scattering, and
the temperature dependence of the normal-state $c$-axis resistivity
are all unknown in these particular materials.

We would like to acknowledge M.B. Ketchen for the design, and M. Bhushan
for the fabrication, of the SQUIDs used in our microscope. We would also 
like
to acknowledge useful conversations with P.W. Anderson,
S. Chakravarty, A.J. Leggett, C. Panagopoulos, and D. van der Marel.

\begin{figure}
\caption{
Scanning SQUID microscope image of a 256$\times$512 micron area of the 
edge of
a single crystal of Hg-1201, cooled in a field of about 10mG, and imaged
at 4.2K. The false color lookup table corresponds to a total variation in
flux through the SQUID pickup loop of 0.55$\Phi_0$. The dashed lines 
indicate
the top and bottom $ab$ faces of the crystal. The box indicates the area
that is expanded in Figure 2.
}

\vspace{0.3in}
\caption{
Expanded view of a 54$\times$100 micron area of the image of Figure 1. The
dashed lines indicate the paths of cross-sections through the data parallel
to the planes displayed in Figure 3.
}

\vspace{0.3in}
\caption{
The symbols are cross-sectional data through the 3 representative vortices
indicated in Figure 2, offset vertically for clarity. The solid lines are
fits to the data, as described in the text. The fit penetration depths
$\lambda_c$ are as labelled in the figure, and the
effective heights are
z$_0$ = 0.6$\pm0.1 \mu$ m(A), 0.8$\pm 0.1 \mu$ m(B), and 0.8$\pm 0.2 \mu 
$m(C).
The error bars are assigned using a doubling of the best-fit $\chi^2$
as a criterion.
}

\vspace{0.3in}
\caption{
(a) $c$-axis resistivity vs temperature for two of our crystals. The 
contact
geometry used is diagrammed in an insert.
(b) Correlation plot between $c$-axis conductivity and $c$-axis penetration
depth, following Basov {\it et al.} (Ref. 29). We have included our 
present 
data for
Hg-1201 and previous data on Tl-1201. The dashed line is the prediction
for diffuse tunneling between superconducting layers, assuming a
gap value 2$\Delta$=28meV. In this figure 124=YBa$_2$Cu$_4$O$_8$,
2212=Bi$_2$Sr$_2$CaCu$_2$O$_8$, ps2=ps3=Pb$_2$Sr$_2$RCu$_3$O$_8$,
214=La$_{1.84}$Sr$_{0.16}$CuO$_4$,123=YBa$_2$Cu$_3$O$_x$.
}

\label{autonum}
\end{figure}

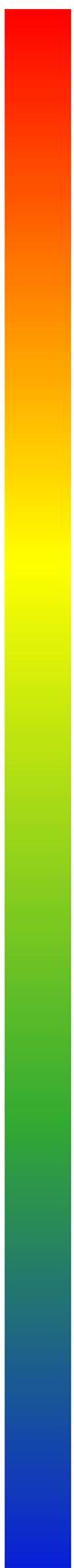
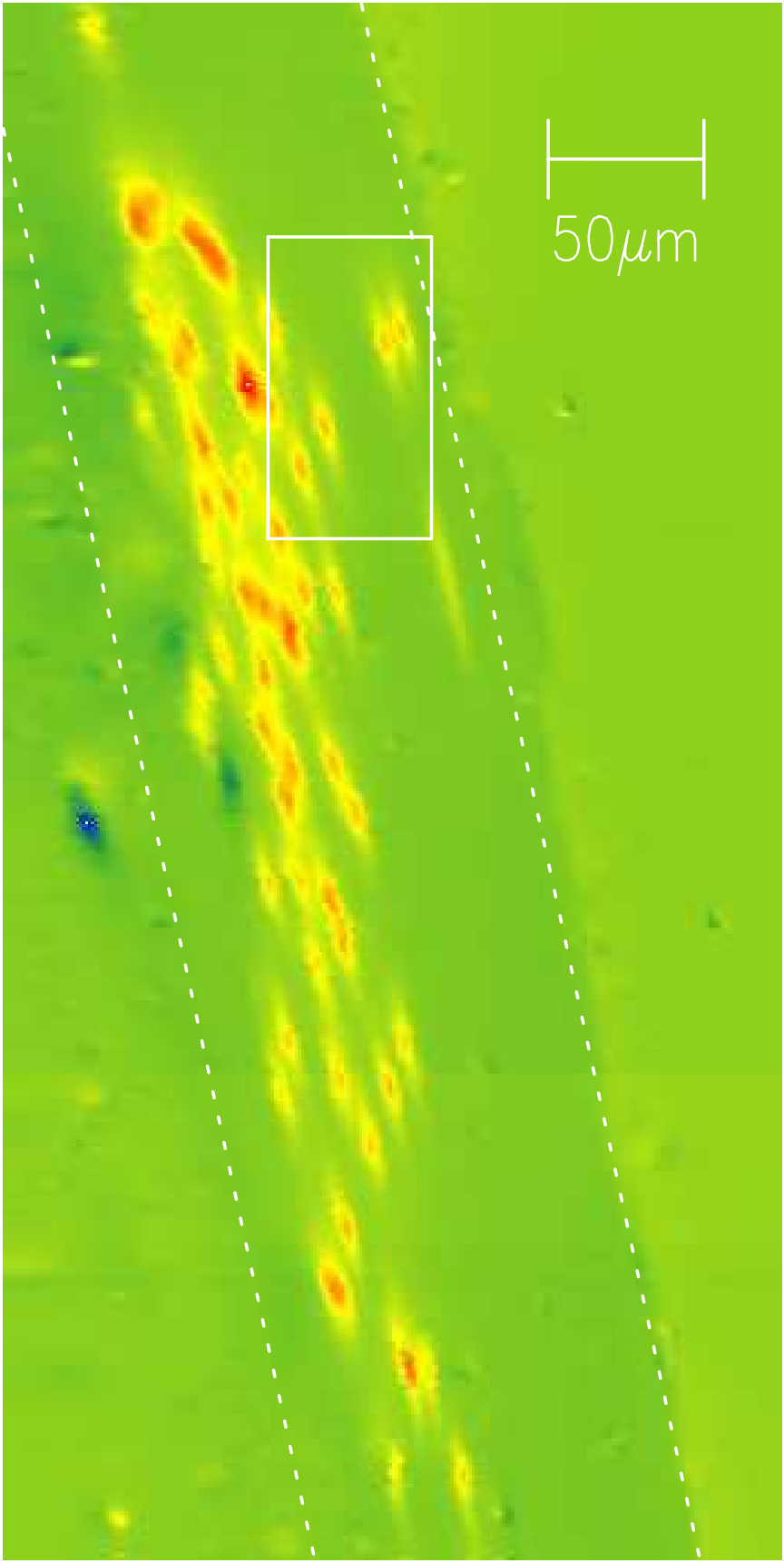
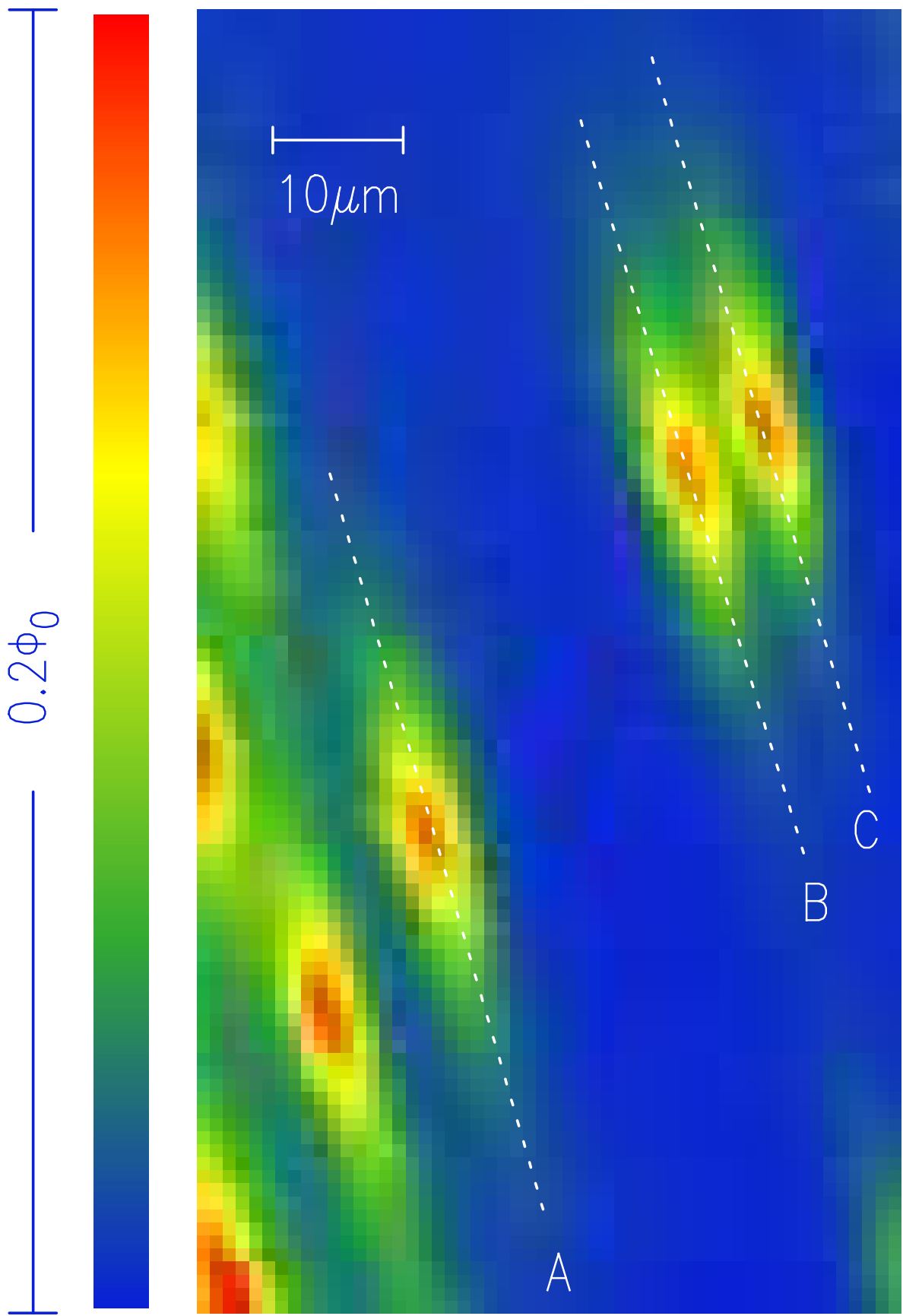
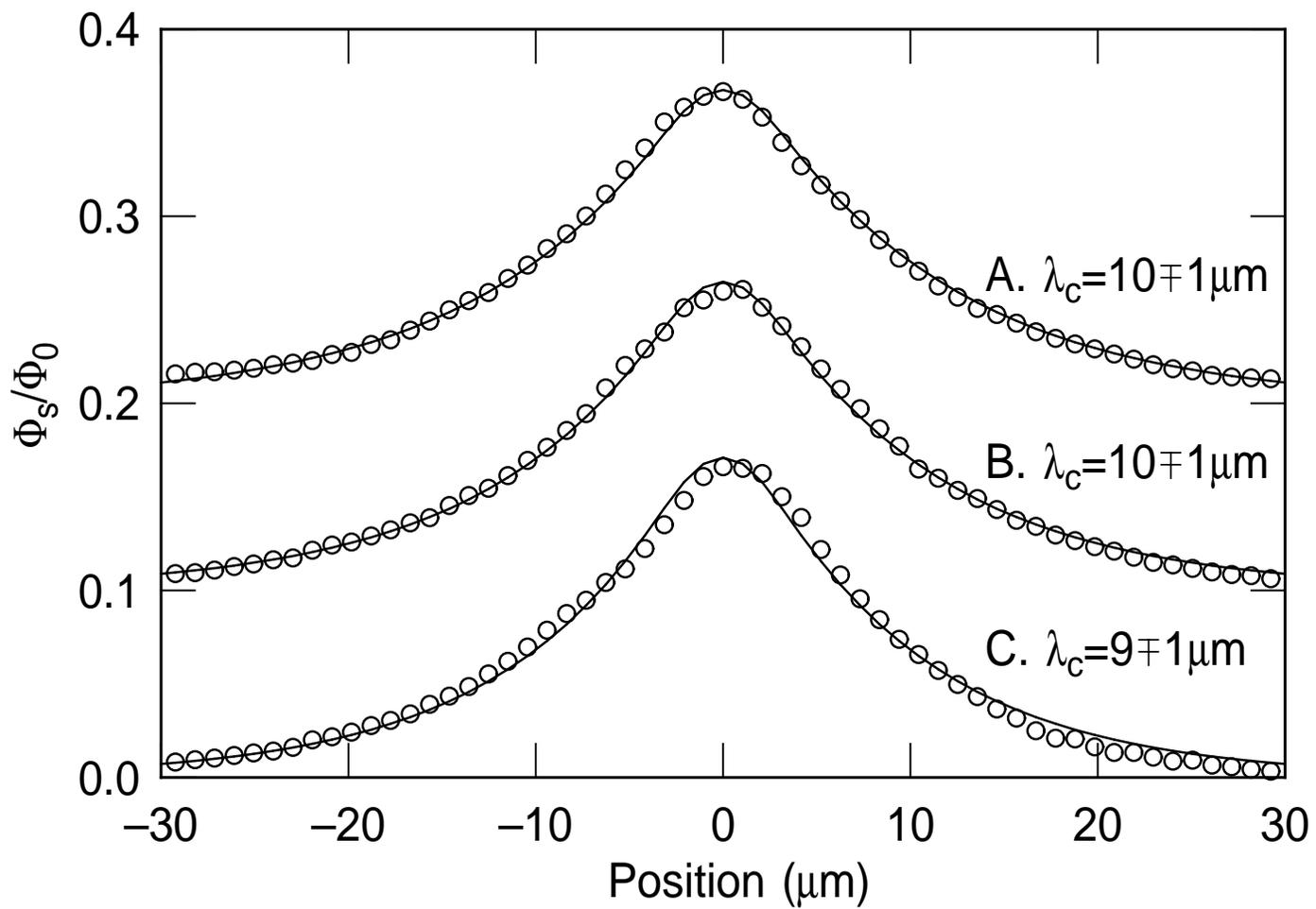
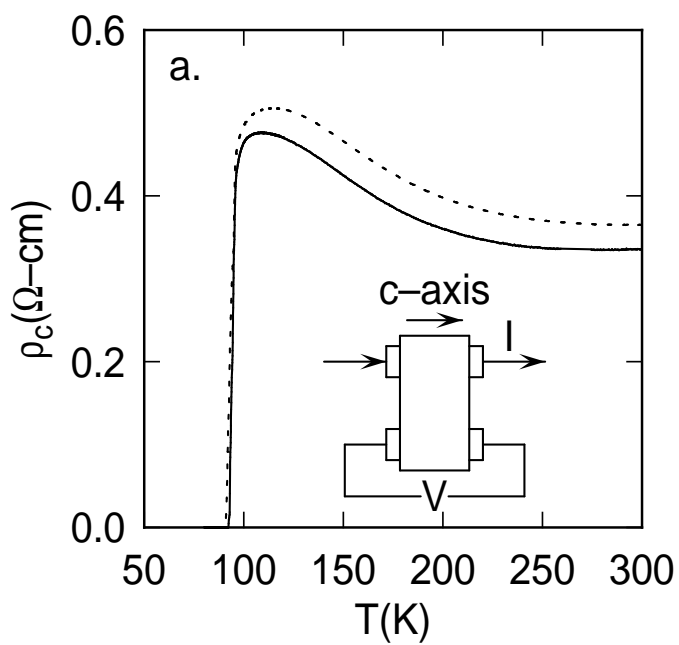
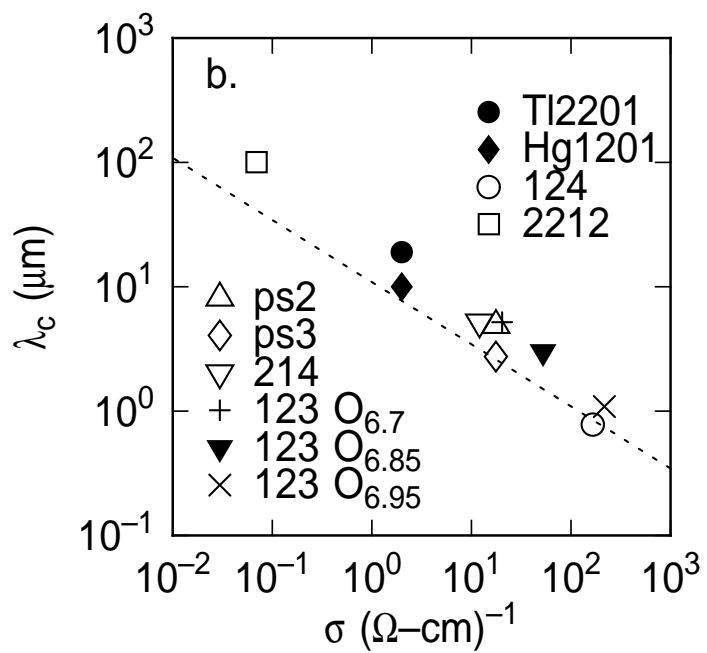
\end{document}